\documentclass[11pt,a4paper]{article}
\usepackage[]{inputenc,graphicx,amsmath,textcomp,amssymb,appendix}
\usepackage{subfigure}  % use for side-by-side figures
\newcommand{\be}{\begin{equation}}
\newcommand{\ee}{\end{equation}}
\newcommand{\bea}{\begin{eqnarray}}
\newcommand{\eea}{\end{eqnarray}}

\catcode`@=12

\begin{document}
\begin{flushright}
%SINP-APC-13/02
\end{flushright}
\thispagestyle{empty}
\begin{center}
{\Large\bf 
{ Inert Doublet Dark Matter with an additional scalar singlet and 125 GeV Higgs Boson }}\\
\vspace{1cm}
{{\bf Amit Dutta Banik} \footnote{email: amit.duttabanik@saha.ac.in}, 
{\bf Debasish Majumdar} \footnote{email: debasish.majumdar@saha.ac.in}}\\
\vspace{0.25cm}
{\normalsize \it Astroparticle Physics and Cosmology Division,}\\
{\normalsize \it Saha Institute of Nuclear Physics,} \\
{\normalsize \it 1/AF Bidhannagar, Kolkata 700064, India}\\
\vspace{1cm}
%%%%%%%%%%%%%%%%%%%%%%%%%%%%%%%%%%%%%%%%%%%%%%%%%%
%{\bf ABSTRACT}
%%%%%%%%%%%%%%%%%%%%%%%%%%%%%%%%%%%%%%%%%%%%%%%%%%
\end{center}
\begin{abstract}
In this work we consider a model for particle dark matter where an extra 
inert Higgs doublet and an additional scalar singlet is added to the Standard 
Model (SM) Lagrangian. The dark matter candidate is obtained from only the 
inert doublet. The stability of this one component dark matter is 
ensured by imposing a $Z_2$ symmetry on this additional inert doublet.
The additional singlet scalar has a vacuum expectation value (VEV) and mixes with the Standard Model 
Higgs doublet resulting in two CP even scalars $h_1$ and $h_2$. 
We treat one of these 
scalars, $h_1$, to be consistent with the SM Higgs like boson of mass around
125 GeV reported by the LHC experiment. These two CP even scalars 
affect the annihilation cross-section of this inert doublet dark matter 
resulting in a larger dark matter mass region that satisfies 
the observed relic density. We also investigate the 
$h_1 \rightarrow \gamma\gamma$ and $h_1 \rightarrow \gamma Z$  processes and 
compared these with LHC results. This is also used to constrain the dark matter
parameter space in the present model. We find that the dark matter candidate 
in the mass region $\frac {m_1} {2} < m_H < m_W$  GeV ($m_1 = 125$ GeV, 
mass of $h_1$) satisfies the recent 
bound from LUX direct detection experiment.   
\end{abstract}
\newpage
\section{Introduction} 
Existence of a newly found Higgs-like scalar boson of mass about 125 GeV
has been reported by recent LHC results.  ATLAS \cite{atlas} and CMS \cite{cms}
independently confirmed the discovery of a new scalars and measured signal
strengths of the Higgs-like scalar to various decay channels separately. ATLAS 
has reported a Higgs to di-photon signal strength ($R_{\gamma \gamma}$) about 
$1.65^{+0.34}_{-0.30}$ \cite{atlas1}. On the other hand Higgs to 
di-photon signal strength 
evaluated by CMS experiment is found to be about $0.78^{+0.28}_{-0.26}$
\cite{cms1}.
Despite the
success of Standard Model (SM) of particle physics, it fails to produce
a plausible explanation of dark matter (DM) in modern cosmology.
Existence of dark matter is firmly established by the observations 
of galaxy rotation
curves and analysis of cosmic microwave background (CMB) etc.
DM relic density  predicted by the PLANCK \cite{planck} and 
WMAP \cite{wmap} results
suggest that about $26.5\%$ of our Universe is constituted by DM. The particle
constituent of dark matter is still unknown and SM of particle physics 
appears inadequate to address the issues regarding dark matter.
The observed dark matter relic density reported by CMB anisotropy 
probes suggests that 
weakly interacting massive particle or WIMP \cite{griest,hooper}.
or WIMP can be assumed to serve as a feasible candidate for dark matter.
Thus, in order to propose a feasible candidate for dark matter
one sould invoke a theory beyond SM and in this regard simple
extension of SM scalar or fermion sector or both could be of interest
in respect of addressing the problem of a viable candidate of 
dark matter and dark matter physics. 
There are other theories though beyond Standard Model (BSM) such as
the elegant theory of Supersymmetry (SUSY) in which the dark matter candidate 
is supposedly the LSP or lightest SUSY particle formed by the superposition 
of neutral gauge bosons and Higgs boson \cite{jungman}.
Extra dimension models \cite{kk} providing Kaluza-Klein dark matter 
candidates are also explored at length in literature. 
Comprehensive studies on simplest extension
of SM with additional scalar singlet where a discrete $Z_2$ symmetry
stabilizes the scalar is studied elaborately in earlier works such
as \cite{zee}-\cite{biswas}.
It is also demonstrated by previous authors that singlet fermion 
extension of SM can also 
be a viable candidate of dark matter \cite{kim}-\cite{sf3}.
SM extensions with two Higgs doublets and a singlet are also addressed
earlier where the additional singlet is 
the proposed of dark matter candidate \cite{aoki,cai}. Among 
various extensions of SM,
a simplest model is to introduce an additional SU(2) scalar 
doublet which produces no VEV. The
resulting model namely 
Inert Doublet Model (IDM) provides a viable explanation for DM.
Satbility of this inert doublet ensured by a discrete $Z_2$ symmetry 
and the lightest inert particle (LIP)
in this theory can be assumed to be a plausible DM
candidate. Phenomenology of IDM has been elaborately studied in literatures
\cite{ma}-\cite{borah}. In the present work, we consider a two Higgs doublet
model (THDM) with an additional scalar singlet, where one of the Higgs
doublet is identical to the inert doublet, i.e., one of the doublet assumes no
VEV and all the SM sector including
the newly added singlet are even under an imposed discrete symmetry $(Z_2)$
while the inert doublet is odd uner that $Z_2$ symmetry. Since inert scalars
do not interact with SM particles, are stable and    
LIP is considered as a potential DM candidate. 
Presence of an additional singlet
scalar enriches the phenomenology of Higgs sector and DM sector.
 
Various ongoing direct detection experiments such as XENON100 \cite{xenon12}, 
LUX \cite{lux}, CDMS \cite{cdms} etc. provide
upper limits on dark matter-nucleon scattering cross sections for 
different possible dark matter mass.
The CDMS \cite{cdms} experiment also claimed to have observed 
three potential signals of dark matter at
low mass region ($\sim 8$ GeV). Direct detection experiments such as
DAMA \cite{dama}, CoGeNT \cite{cogent} and
CRESST \cite{cresst} also provide bounds on dark matter-nucleon
scattering cross sections for different dark matter masses. 
These experiments predict
the presence of low mass dark matter candidate that contradict with XENON100 or LUX results
since they provide a much lower bound for 
dark matter-nucleon scattering cross section. We also test our 
model by calculating the
$R_{\gamma \gamma}$ for $h\rightarrow \gamma \gamma$ signal 
and comparing the same 
with those given by LHC experiment.
 
In this work, we
consider an Inert Doublet Model (IDM) along with an additional singlet scalar 
field $S$. A discrete $Z_2$ symmetry, under which all SM particles
along with the singlet scalar $S$ are even while the inert doublet considered is odd, allows
the LIP $(H)$ to remain stable and serve as a viable dark 
matter candidate. Additional scalar singlet having a non zero
VEV mixes with the SM Higgs, provides two CP even Higgs 
states. 
We consider one of the scalars, $h_1$, to be the SM-like Higgs. Then 
$h_1$ should be compatible with SM Higgs and one can compare the 
relevant calculations for $h_1$ with that obtained in LHC experiment.
The model parameter space is first constrained by theoretical conditions
such as vacuum stability, perturbativity, unitarity and then by 
the relic density bound given 
by PLANCK/WMAP experiments. We evaluate the direct detection scattering 
cross-section $\sigma_{\rm SI}$ with the resulting constrained parameters 
for different LIP masses $m_H$ and investigate the regions in 
$\sigma_{\rm SI} - m_H$ plane that satisfy the bounds from experiments 
like LUX, XENON etc.  We also 
calculate the signal strength $R_{\gamma\gamma}$ for 
$h_1 \rightarrow \gamma\gamma$ channel in the present framework
and compare them with the experimentally obtained limits  
for this quantity from CMS and ATLAS experiments. 
This will further constrain the model parameter space. We thus obtain  
regions in $\sigma_{\rm SI} - m_H$ plane in the present framework that 
satisfy not only the experimental results for dark matter relic 
density and scattering 
cross-sections but compatible with LHC results too.     

The paper is organised as follows. In Sec.~\ref{S:modelandbounds} 
we present a description of
the model and model parameters with relevant bounds from theory 
(vacuum stability, pertubativity and unitarity)  and experiments (PLANCK/WMAP, 
direct detection experiments, LHC etc.). In Sec.~\ref{S:darkmatter} 
we describe the relic density and annihilation cross section measurements for
dark matter and modified $R_{\gamma \gamma}$ and $R_ 
{\gamma Z}$ processes due to inert charged scalars.
We constrain the model parameter space satisfying the relic density
requirements  of dark
matter and present the correlation between $R_{\gamma \gamma}$ and
$R_{\gamma Z}$ processes in Section \ref{S:Analysis}.
In Sec.~\ref{S:ddetection}, we further 
constrain the results by direct detection bounds on dark matter. Finally,
in Sec.~\ref{S:summary} we summarize the work briefly with concluding remarks.  

\section{The Model}
\label{S:modelandbounds}
\subsection{Scalar Sector}
\label{ss:model}
In our model we add an additional SU(2) scalar doublet and a 
%We consider Inert Doublet Model (IDM) in addition with 
a real scalar singlet
$S$. Similar to the widely studied inert doublet model or IDM where the added 
SU(2) scalar doublet to the SM Lagrangian is made ``inert" (by imposing 
a $Z_2$ symmetry that ensures no interaction with SM fermions and 
the inert doublet does not generate any vev), here too 
the extra doublet is assumed to be odd under a discrete $Z_2$ symmetry 
(IDM). Under this $Z_2$ symmetry however, 
all SM particles as also the added singlet $S$ remain unchanged. 
The potential is expressed as
\bea
V &=& m_{11}^2 {\Phi_1}^\dagger {\Phi_1} + m_{22}^2  {\Phi_2}^\dagger {\Phi_2} + {1 \over 2}
m_s^2 S^2 + \lambda_1 ({\Phi_1}^\dagger {\Phi_1})^2
+ \lambda_2 ({\Phi_2}^\dagger {\Phi_2})^2 + \lambda_3
(\Phi_1^\dagger \Phi_1)(\Phi_2^\dagger \Phi_2) \nonumber \\ &+&
\lambda_4 (\Phi_2^\dagger \Phi_1)(\Phi_1^\dagger \Phi_2) + {1 \over 2} \lambda_5
[(\Phi_2^\dagger \Phi_1)^2 + (\Phi_1^\dagger \Phi_2)^2] + \rho_1
(\Phi_1^\dagger \Phi_1) S + \rho'_1 (\Phi_2^\dagger \Phi_2) S \nonumber
\\  &+&  \rho_2 S^2 (\Phi_1^\dagger \Phi_1) +  \rho'_2
S^2 (\Phi_2^\dagger \Phi_2) +{1 \over 3} \rho_3 S^3 + {1 \over 4} \rho_4 S^4 ,
\label{1}
\eea
where $m_k (k=11,22,s)$ etc. and all the coupling parameters 
($\lambda_i$, $\rho_i$, ${\rho'}_i$, $i=1,2,3,...$ etc.) 
are assumed to be real.
In Eq. \ref{1}, $\Phi_1$ is the ordinary SM Higgs doublet and $\Phi_2$ is the
inert Higgs doublet. After spontaneous symmetry breaking $\Phi_1$ 
and $S$ acquires VEV and expressed as
\bea
~~~~~~~~~~~\Phi_1 =\left( \begin{array}{c}
                           0  \\
        \frac{1}{\sqrt{2}}(v+h)  
                 \end{array}  \right) \, ,                     
~~~~~~~~~~~\Phi_2 =\left( \begin{array}{c}
                           H^+   \\
        \frac{1}{\sqrt{2}}(H+iA)  
                 \end{array}  \right) \, ,
~~~~~~~~~~~~~S = v_s + s \, . 
\label{2}
\eea  
In the above $v_s$ denotes the VEV of the field $S$ and $s$ is the real
singlet scalar.
Relation among model parameters can be obtained from
the extremum conditions of the potenial expressed in Eq.~\ref{1} and are given as
\bea
m_{11}^2+\lambda_1 v^2+ \rho_1 v_s + \rho_2 v_s^2 = 0 \, ,\nonumber \\ 
m_s^2+ \rho_3 v_s + \rho_4 v_s^2 + \frac{\rho_1 v^2}{2v_s} + \rho_2 v^2 =0 \, .\nonumber 
\eea 
Mass terms of various scalar particles as derived from the potential are
\bea
\mu_h^2&=&2 \lambda_1 v^2 \nonumber \\
\mu_s^2&=&\rho_3 v_s + 2\rho_4 v_s^2 - \frac{\rho_1 v^2}{2 v_s} \nonumber \\
\mu_{hs}^2&=&(\rho_1+ 2 \rho_2 v_s) v  \nonumber \\
m_{H^{\pm}}^{2}&=&m_{22}^{2}+\lambda_{3}\frac{v^{2}}{2}+\rho'_1 v_s+\rho'_2 v_s^2  \nonumber \\
m_{H}^{2}&=&m_{22}^{2}+(\lambda_{3}+\lambda_{4}+\lambda_{5})\frac{v^{2}}{2}+\rho'_1 v_s+\rho'_2 v_s^2 \nonumber \\
m_{A}^{2}&=&m_{22}^{2}+(\lambda_{3}+\lambda_{4}-\lambda_{5})\frac{v^{2}}{2}
+\rho'_1 v_s+\rho'_2 v_s^2\,\, . 
\label{3}   
\eea 
The mass eigenstates $h_1$ and $h_2$ are linear combinations of $h$ and $s$ and can be 
written as
\bea h_1=s~\sin\alpha+h~\cos\alpha  \, , \nonumber \\
h_2=s~\cos\alpha-h~\sin\alpha \, , 
\label{4}
\eea   
$\alpha$ being the mixing angle between $h_1$ and $h_2$, is given by
\be \tan\alpha\equiv\frac{x}{1+\sqrt{1+x^2}}\, , 
\label{5}
\ee
where $x=\frac{2\mu^2_{hs}}{(\mu^2_h-\mu^2_s)}$.
Masses of the physical neutral scalars $h_1$ and $h_2$ are 
\be
m^2_{1,2}=\frac{\mu_h^2+\mu_s^2}{2}\pm\frac{\mu_h^2-\mu_s^2}{2}\sqrt{1+x^2}.
\label{6}
\ee 
We consider $h_1$ with 
mass $m_1=125$ GeV as the SM-like Higgs boson and the mass of the other scalar
$h_2$ in the model is denoted as $m_2$ with  $m_2>m_1$.
Couplings of the physical scalars $h_1$ and $h_2$ with SM 
particles are modified by the factors 
$\cos \alpha$ and $\sin \alpha$ respectively.
In the present framework $H$ and $A$ are stable as long 
as the $Z_2$ symmetry is unbroken and hence these
neutral scalars can be viable candidates for dark matter. 
Here, the coupling $\lambda _{5}$ serves as a mass splitting factor between 
$H$ and $A$. We consider $H$ to be the lightest inert particle (LIP) which is
stable and is DM candidate in this work. We take $\lambda _{5} < 0$ 
in order to make $H$ to be the lightest stable inert 
particle. In the present framework, both the scalars $h_1$ and $h_2$ couple
with the lightest inert particle $H$. Couplings of the scalar bosons 
($h_1$ and $h_2$) with the inert dark matter $H$ are given by
\bea
\lambda_{h_1HH}v=\left(\frac{\lambda_{345}}{2}c_{\alpha}-\frac{\lambda_s}{2}s_{\alpha}\right)v \, ,\nonumber \\
\lambda_{h_2HH}v=\left(\frac{\lambda_{345}}{2}s_{\alpha}+\frac{\lambda_s}{2}c_{\alpha}\right)v
\label{7} 
\eea 
where $\lambda_{345} = \lambda_3 + \lambda_4 + \lambda_5 $,
$\lambda_s =\frac{\rho_1'+2 \rho_2' v_s}{v} $ and $s_{\alpha}(c_{\alpha})$ denotes
$\sin\alpha(\cos\alpha)$. Couplings of scalar
bosons with charged scalars $H^{\pm}$ are
\bea
\lambda_{h_1H^+H^-}v=\left(\lambda_{3}c_{\alpha}-{\lambda_s}s_{\alpha}\right)v \, , \nonumber \\
\lambda_{h_2H^+H^-}v=\left(\lambda_{3}s_{\alpha}+{\lambda_s}c_{\alpha}\right)v.
\label{8} 
\eea
\subsection{Constraints}
\label{ss:constraints}
The model parameters are bounded by theoretical and experimental constraints.\\
\begin{itemize}
\item {\bf Vacuum Stability} - 
Vacuum stability constraints requires the potential to remain bounded from
below. Conditions for the stability of the vacuum are 
\bea
\lambda_1,\,\lambda_2,\,\rho_4 &>& 0 \,\, ,
\nonumber \\
\lambda_3 + 2\sqrt{\lambda_1\lambda_2} & > & 0\, ,
\nonumber\\
\lambda_3 +\lambda_4 +\lambda_5+ 2\sqrt{\lambda_1\lambda_2} & > & 0\, ,
\nonumber\\
\rho_2 +\sqrt{\lambda_1\rho_4} & > & 0\, , 
\nonumber \\
\rho_2' +\sqrt{\lambda_2\rho_4} & > & 0\, .
\label{9}
\eea
\item {\bf Pertubativity} - 
For a theory to be acceptable in perturbative limits, we have to 
constrain high energy
quartic interactions at tree level. The eigenvalues $|\Lambda_i|$ of quartic
couplings (scattering) matrix must be smaller than $4\pi$.\\
\item {\bf LEP}
LEP\cite{lep} results constrains the $Z$ boson decay width and masses of scalar 
particles
\bea
m_H + m_A  >  m_Z \,\, , \nonumber \\
m_{H^{\pm}}  >  79.3 ~\rm{GeV}.
\eea 
\item {\bf Relic Density} - 
Parameter space is also constrained by the measurement of relic density of dark
matter candidate. Relic density of the lightest inert particle (LIP) serving  as a 
viable candidate for dark matter in the present model must satisfy PLANCK/WMAP results,
\be
\Omega_{\rm DM} h^2 = 0.1199{\pm 0.0027}\,\, .
\label{wmap,planck}
\ee
\item {\bf Higgs to Diphoton Rate $\bf{R_{\gamma \gamma}}$}
Bound on Higgs to two photon channel has been obtained 
from experiments performed
by LHC. The reported singal strength for the Higgs 
to diphoton channel from ATLAS and CMS are given as
\be
R_{\gamma \gamma}|_{\rm{ATLAS}} = 1.65^{+0.34}_{-0.30} \, , \\ \nonumber
R_{\gamma \gamma}|_{\rm{CMS}}= 0.78^{+0.28}_{-0.26}\,\, .  
\ee
\item {\bf Direct Detection Experiments} - 
The bounds on dark matter from direct detection
experiments are based on the 
elastic scattering of the dark 
matter particle off a scattering nucleus. Dark matter direct detection 
experiments set constraints on the dark matter - nucleus (nucleon) elastic 
scattering cross section. Limits 
on scattering cross sections for different dark matter mass 
cause further restrictions on the model parameters. 
Experiments like CDMS, DAMA, CoGeNT, CRESST etc. provide effective bounds 
on low mass dark matter. Stringent bounds on midddle mass 
and high mass dark matter are
obtained from XENON100 and LUX experiments.
\end{itemize}
\section{Dark matter}
\label{S:darkmatter}
\subsection{Relic density}
\label{ss:relicdensity}
Relic density of dark matter is constrained by the results of PLANCK and WMAP.
Dark matter relic abundance for the model is evaluated by solving the
evolution of Boltzmann equation given as \cite{kolb}
\bea
\frac{{\rm d} n_H}{{\rm d} t} + 3 {\rm H}n_H &=& - \langle \sigma {\rm v} \rangle (n_H^{2}-n_{H\rm{eq}}^{2})\,\, .
\label{12}
\eea
In Eq. \ref{12}, $n_H(n_{Heq})$ denotes the number density (equilibrium 
number density) of dark matter $H$ and $\rm H$ is the Hubble constant.
In Eq. \ref{12}, $\langle \sigma {\rm v} \rangle$ denotes the thermal averaged
annihilation cross section of dark matter particle to SM species. 
The  dark matter relic density can be obtained by solving Eq. \ref{12} and is 
written as 
\bea
\Omega_{\rm{DM}}{\rm h}^2 &=& \frac{1.07\times 10^9 x_F}
{\sqrt{g_*}M_{\rm Pl}\langle \sigma {\rm v} \rangle}\,\, .
\label{13}
\eea  
In the above, $M_{\rm Pl}=1.22\times10^{19}$GeV, is the Planck scale mass 
whereas $g_*$ is the effective number of degrees of freedom in thermal 
equilibrium and 
$\rm h$ is the Hubble parameter in unit 
of $100\,~{\rm km}~{\rm s}^{-1} {\rm Mpc}^{-1}$.
In Eq. \ref{13}, $x_F = M/T_F$, where $T_F$ is the freeze out temperature of the annihilating
particle and $M$ is the mass of the dark matter ($m_H$ for the present scenario).
Freeze out temperature $T_F$ for the dark matter 
is obtained from the iterative solution to the equation
\bea
x_F &=& \ln \left ( \frac{M}{2\pi^3}\sqrt{\frac{45M_{\rm{Pl}}^2}{2g_*x_F}}
\langle \sigma \rm{v} \rangle \right )\,\, . 
\label{14}
\eea   
 
\subsection{Annihilation cross section}
\label{ss:anncross}
Annihilation of inert dark matter $H$ to SM particles is governed by 
processes involving scalar ($h_1,h_2$) mediated s$(\simeq 4m_H^2)$ channels. 
Thermal averaged annihilation 
cross section $\langle \sigma \rm{v}\rangle$ of dark matter $H$
to SM fermions are given as 
\bea
\langle{\sigma {\rm{v}}}_{H H\rightarrow f\bar f}\rangle &=&  n_c\frac{{m^2_f}}{\pi}
\beta_f^{3}
\left|\frac{\lambda_{h_1HH}\cos{\alpha}}{4{m^2_{H}}-{m^2_1}+i\Gamma_1 m_1}
+\frac{\lambda_{h_2HH}\sin{\alpha}}{4{m^2_{H}}-{m^2_2}+i\Gamma_2 m_2}\right|^2 \,\, . 
\label{15}
\eea
In the above, $m_x$ represents mass of the particle $x (\equiv f,~H$ etc.), 
$n_c$ is the colour quantum number (3 for quarks and 1 for leptons) with 
$\beta_a = \sqrt{1-\frac{m_a^2}{m_H^2}}$ and $\Gamma_i(i=1,2)$ denotes 
the total decay 
width of each of the two scalars $h_1$ and $h_2$.
For DM mass $m_H>(m_W,m_Z$), annihilation of DM to gauge boson 
($W$ or $Z$)
channels will yield high annihilation cross-section.
Since $\Omega_{\rm{DM}}\sim \langle \sigma \rm{v} \rangle^{-1}$ (Eq. \ref{13}),
the relic density for the dark matter with mass 
$m_H>m_W~{\rm or}~m_Z$ in the present model in fact falls below the 
relic density
given by WMAP or PLANCK as the four point interaction channel
$HH\rightarrow W^+ W^- {\rm or}~ ZZ$ will be accessible and as a 
result increase in total
annihilation cross-section will be observed.
Thus the possibility of a single component DM in the present
framework is excluded for mass $m_H>m_W,~m_Z$
\footnote{Similar results for IDM are also obtained in previous work 
(Ref. \cite{anirban2})
where a two component dark matter was considered in order to circumvent 
this problem.}. 
Higgs like boson $h_1$ and the scalar $h_2$ may also decay
to dark matter candidate $H$ when the condition 
$m_H < m_i/2(i=1,2)$ is satisfied.
Contributions of invisible decay widths of $h_1$ and $h_2$ are taken 
into account
when the condition $m_H < m_i/2(i=1,2)$ is satisfied. Invisible decay width is 
represented by the relation
\be
\Gamma_{inv}(h_i \rightarrow 2H)= \frac{\lambda^2_{h_iHH} v^2}{16\pi m_i}\sqrt{1-\frac{4m^2_H}{m^2_i}}\, . 
\label{16}
\ee
\subsection{Modification of $R_{\gamma \gamma}$ and $R_{\gamma Z}$}
\label{ss:decaywidth}
Recent studies of IDM \cite{arhrib,maria,goudelis} and two Higgs doublet models 
\cite{posch,dipankar} have reported that a low mass charged scalar could
possibly enhance the $h_1 \rightarrow \gamma \gamma$ signal strength $R_{\gamma
\gamma}$. Correlation of $R_{\gamma \gamma}$ with $R_{\gamma Z}$ is also 
accounted for as well \cite{maria,dipankar}. The quantities $R_{\gamma \gamma}$ and 
$R_{\gamma Z}$ are expressed as
\bea
R_{\gamma \gamma} = \frac{\sigma(pp\rightarrow h_1)}
{\sigma(pp\rightarrow h)^{\rm SM}}
\frac{Br(h_1\rightarrow \gamma \gamma)}{Br(h\rightarrow \gamma \gamma)^{\rm SM}}
\label{17}
\eea
\bea
R_{\gamma Z} = \frac{\sigma(pp\rightarrow h_1)}
{\sigma(pp\rightarrow h)^{\rm SM}}
\frac{Br(h_1\rightarrow \gamma Z)}{Br(h\rightarrow \gamma Z)^{\rm SM}}\,\, ,
\label{18}
\eea
where $\sigma$ is the Higgs production cross section and $Br$ represents the 
branching ratio of Higgs to final states.
Branching ratio to any final state is given by the ratio of
partial decay width for the particular channel to the total
decay width of decaying particle. For IDM with 
additional singlet scalar, the ratio 
$\frac{\sigma(pp\rightarrow h_1)}{\sigma(pp\rightarrow h)^{\rm SM}}$
in Eqs. \ref{17}-\ref{18} is represented by a factor $\cos^2 \alpha$.
Standard Model branching ratios $Br(h\rightarrow \gamma \gamma)^{\rm SM}$
and $Br(h\rightarrow \gamma Z)^{\rm SM}$ for a 125 GeV Higgs boson is 
$2.28\times10^{-3}$ and $1.54\times10^{-3}$ respectively \cite{denner}. 
To evaluate the branching ratios $Br(h_1\rightarrow \gamma \gamma)$
and $Br(h_1\rightarrow \gamma Z)$, we compute the total decay width
of $h_1$. Invisible decay of $h_1$ to dark matter particle $H$ is also
taken into account and evaluated using Eq. \ref{16} when the condition
$m_H < m_1/2$ is satisfied. Partial decay widths $\Gamma(h_1 \rightarrow
\gamma \gamma)$ and $\Gamma(h_1 \rightarrow \gamma Z)$ according to the model
are given as
\bea
\Gamma(h_1\rightarrow \gamma\gamma)&=&\frac{G_F\alpha_s^2m_1^3}{128\sqrt{2}\pi^3}
\left |\cos\alpha\left(\frac{4}{3}  F_{1/2}\left(\frac{4m_t^2}{m_1^2}\right)
+ F_1 \left(\frac{4m_W^2}{m_1^2} \right)\right)
+\frac{\lambda_{h_1H^+H^-}v^2}{2m_{H^{\pm}}^2}
F_0 \left(\frac{4m_{H^{\pm}}^2}{m_1^2}\right)
\right |^2, \nonumber \\
\Gamma(h_1\rightarrow \gamma Z) &= &\frac{G_F^2\alpha_s}{64\pi^4} m_W^2 m_1^3 \left(1-\frac{m_Z^2}{m_1^2}\right)^3
\left|-2\cos\alpha \frac{1-\frac{8}{3}s^2_W}{c_W}
F_{1/2}'\left(\frac{4m_t^2}{m_1^2},\frac{4m_t^2}{m_Z^2}\right) \right. \nonumber \\
&& \left .
-\cos\alpha F_1'\left(\frac{4m_W^2}{m_1^2},\frac{4m_W^2}{m_Z^2}\right)
+\frac{\lambda_{h_1H^+H^-}v^2}{2m_{H^{\pm}}^2}
\frac{(1-2s^2_W)}{c_W}I_1\left(\frac{4m_{H^{\pm}}^2}{m_1^2},\frac{4m_{H^{\pm}}^2}{m_Z^2}\right)
\right|^2,
\label{19}
\eea
where $G_F$ is the Fermi constant, $m_x$ denotes the mass of particle 
$x(x\equiv 1,W,Z,t,H^{\pm})$ etc. and $s_W(c_W)$ represents 
$\sin \theta_W(\cos \theta_W)$,
$\theta_W$ being the weak mixing angle. Expressions for various 
loop factors  
($F_{1/2},~F_1,~F_0,~F_{1/2}',~F_1'$ and $I_1$) appeared in Eq. \ref{19}
are given 
in Appendix A. It is to be noted that a similar derivation of decay widths
and signal strengths $({R'}_{\gamma \gamma}~\rm{or}~{R'}_{\gamma Z})$ 
for the other scalar $h_2$ can be obtained by replacing 
$m _1,~\cos{\alpha},~\lambda_{h_1H^+H^-}$ 
with $m _2,~\sin{\alpha},~\lambda_{h_2H^+H^-}$ respectively and this is  
addressed in Sec. \ref{S:ddetection}. 
\section{Analysis of $R_{\gamma \gamma}$ and $R_{\gamma Z}$}
\label{S:Analysis}
In this section we compute the  quantities $R_{\gamma \gamma}$ and 
$R_{\gamma Z}$ in the framework of the present model.
We restrict the allowed model parameter space for our analysis using the 
vacuum stability, perturbative unitarity, LEP bounds 
along with the relic density constraints described in Section 2.2.
Dark matter relic density is evaluated by solving
the Boltzmann equation 
presented in Section 3.1 with the expression for annihilation cross section
given in Eq. \ref{15}.
Model parameters $(\lambda_i,\rho_i)$, should remain small in order to satisfy 
perturbative bounds and relic density constraints. 
Calculations are made for the model parameter limits given below,
\bea
m_1 = 125 ~\rm{GeV}\, , \nonumber \\
80 ~{\rm{GeV}} \leq m_{H^{\pm}} \leq 400~{\rm{GeV}}\, , \nonumber \\
0 < m_H < m_{H^{\pm}},~m_A\, , \nonumber \\
0 < \alpha < \pi/2\, ,\nonumber \\
-3 \leq \lambda_3 \leq 3\, , \nonumber \\
-3 \leq \lambda_{345} \leq 3\, , \nonumber \\
-3 \leq \lambda_s \leq 3\, .
\label{20} 
\eea 
The couplings $\lambda_{h_1HH}$ 
and $\lambda_{h_2HH}$ (Eq. \ref{7})
are required to calculate the scattering cross-section of the dark matter
off a target nucleon. Dark matter direct detection experiments are based
on this scattering processes whereby the recoil energy of the scattered
nucleon is measured. Thus the couplings $\lambda_{h_1HH}$ and $\lambda_{h_2HH}$
can be constrained by comparing the computed values of the scattering
cross-section for different dark matter masses with those given by different
dark matter direct detection experiments. 
In the present work, $|\lambda_{h_1HH},\lambda_{h_1H^+H^-}|\leq 3$ is
adopted. The 
following bounds on parameters will also constrain the couplings
$\lambda_{h_2HH}$ and $\lambda_{h_2H^+H^-}$ (Eqs. \ref{7}-\ref{8}). 
Using Eqs. \ref{12}-\ref{16} we scan over the parameter space
mentioned 
in Eq. \ref{20} where we also impose the conditions 
$|\lambda_{h_1H^+H^-},~\lambda_{h_1HH}|\leq 3$ to calculate the relic 
densities for the LIP
dark matter in the present model. Comparison with the experimentally 
obtained range of
dark matter relic density with the calculated values restricts 
the allowed model parameter
space and gives the range of mass that satisfies observed DM relic density. 
We have made our calculations for three diffrent values of singlet scalar $(h_2)$ 
mass namely
$m_2=140,~150$ and $160$ GeV. Scanning of the full parameter space yields
that for all the cases considered, the limits $-2.1 \leq \lambda_{h_1HH} \leq 1.5$ and
$|\lambda_{h_2HH}|\leq 2.0$ are required for satisfying observed DM relic abundance.
The condition $|\lambda_{h_1H^+H^-}|\leq 3$ also bounds the coupling 
$\lambda_{h_2H^+H^-}$. Our calculation reveals that $|\lambda_{h_2H^+H^-}|\leq 4$ 
is needed in order to satify observed relic density of dark matter. Using the 
allowed parameter space thus obtained, we measured the 
signal strengths $R_{\gamma \gamma}$ and $R_{\gamma Z}$   
(Eqs. \ref{17}-\ref{18}) by evaluating the corresponding decay widths given in Eq. \ref{19}. 
\begin{figure}[h!]
\centering
\subfigure[]{
\includegraphics[scale= 0.65,angle=0]{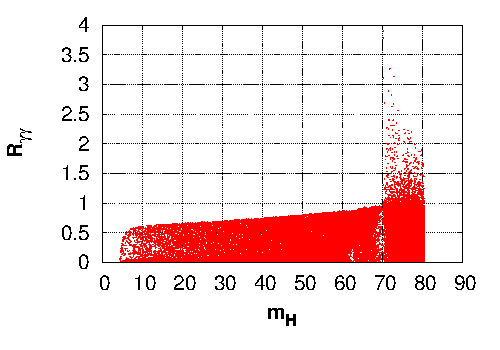}}
\subfigure []{
\includegraphics[scale= 0.65,angle=0]{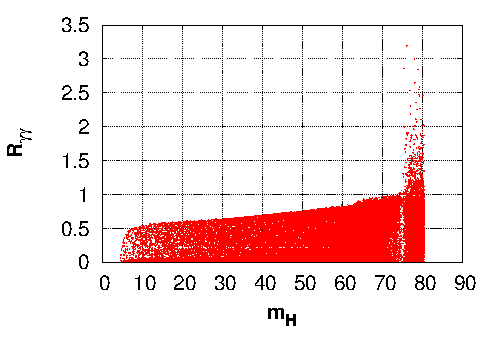}}
\subfigure []{
\includegraphics[scale= 0.65,angle=0]{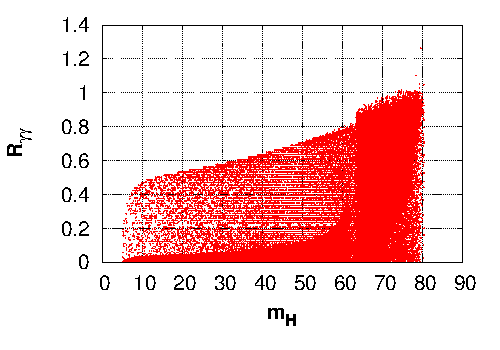}}
\caption{Variation of $R_{\gamma \gamma}$ with DM mass $m_H$ satifying DM relic density 
for $m_2 = 140,150~\rm{and}~160$ GeV.}
\label{fig1}
\end{figure}
\par
In Fig. \ref{fig1}(a-c), shown are the regions in the  $R_{\gamma \gamma}-m_H$ 
plane for the 
parameter values that satisfy 
DM relic abundance. As mentioned earler,
results are presented for three values of $h_2$
mass namely $140,150$ and $160$ GeV. 
Since for low mass DM region, invisible decay channel of
$h_1$ to DM pair remains open, enhancement of $R_{\gamma \gamma}$ is not 
possible in this regime. 
$R_{\gamma \gamma}$ becomes greater than unity
near the region where $m_H \gtrsim m_2/2$.
The region that describe the 
$R_{\gamma \gamma}$ enhancement is reduced with increasing
$h_2$ mass and thus enhacement is not favoured for higher values of $h_2$
mass. 
For the rest of the 
allowed DM mass parameter space, $R_{\gamma \gamma}$ remains less than 1. The
results presented in Fig. \ref{fig1} indicate that observed enhancement of the 
$h_1\rightarrow \gamma \gamma$ signal could be a possible 
indication of the presence of $h_2$ since $R_{\gamma \gamma} 
\gtrsim 1$ occurs near the resonance of $h_2$. The $R_{\gamma \gamma}$ value
depends on the coupling $\lambda_{h_1 H^+H^-}$ and becomes greater than unity
only for $\lambda_{h_1 H^+H^-} < 0 $ and interfers constructively 
with the other loop contributions. Technically,
$R_{\gamma \gamma}$ depends on the values of $h_2$ mass, charged scalar mass $m_{H^{\pm}}$, coupling 
$\lambda_{h_1 H^+H^-}$ and the decay width of invisible decay channel 
($\Gamma_{inv}(h_1\rightarrow H H)$). 
A similar variation for the $h_1\rightarrow \gamma Z$ channel
(computed using Eqs. \ref{18}-\ref{19} and Eq. \ref{20}) yields
lesser enhancement for $R_{\gamma Z}$ in comparison with $R_{\gamma \gamma}$.
This phenomenon can also be verified from the 
correlation between $R_{\gamma \gamma}$ and $R_{\gamma Z}$.
\begin{figure}[h!]
\centering
\subfigure[]{
\includegraphics[scale= 0.65,angle=0]{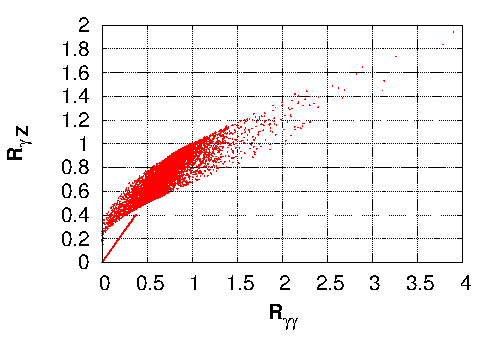}}
\subfigure []{
\includegraphics[scale= 0.65,angle=0]{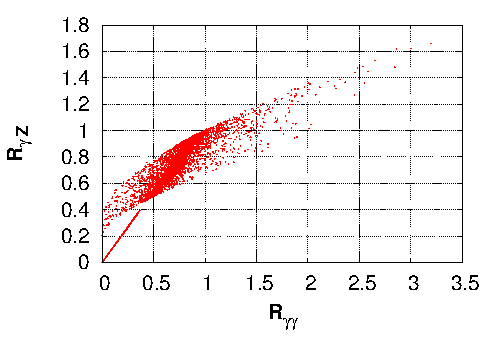}}
\subfigure []{
\includegraphics[scale= 0.65,angle=0]{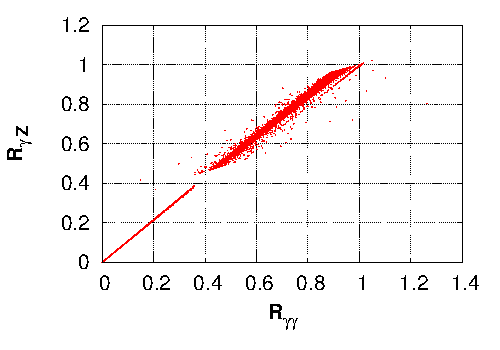}}
\caption{Correlation plots between $R_{\gamma \gamma}$ and $R_{\gamma Z}$ for three choices of $h_2$ mass
($140,150~\rm{and}~160$ GeV).}
\label{fig2}
\end{figure}
The correlation between the signals $R_{\gamma \gamma}$ and $R_{\gamma Z}$
is shown in Fig. \ref{fig2}a - Fig. \ref{fig2}c for 
$m_2 = 140,150,160$ GeV respectively.  
Variations of $R_{\gamma \gamma}$ and $R_{\gamma Z}$
satisfy all necessary parameter constraints taken into account 
inclusive of the relic requirements for DM. In this case (Fig. \ref{fig2}),
we further constrain the parameter space of $\alpha$ 
mentioned in Eq. \ref{20} by
imposing the condition $0 < \alpha < \pi/4$.
This condition ensures that $h_1$ is the SM-like Higgs boson \cite{kim,sf3}.  
Fig. \ref{fig2} also indicates that, with increase
in the mass ($m_2$) of $h_2$, enhancement of $R_{\gamma \gamma}$ 
and $R_{\gamma Z}$ 
are likely to reduce. For $m_2=140$ GeV, $R_{\gamma \gamma}$ 
enhances up to
four times whereas $R_{\gamma Z}$ increases 
nearly by a factor 2 with respect to corresponding values predicted SM.
On the other hand, for $m_2=160$ GeV,
$R_{\gamma \gamma}$ varies linearly 
with $R_{\gamma Z}$ $(R_{\gamma \gamma}\simeq
R_{\gamma Z})$ without any significant enhacement.
In addition Fig. \ref{fig2} suggests that 
for $\cos\alpha \gtrsim 1/\sqrt{2}$, a considerable
portion of allowed parameter space with lower values of $R_{\gamma \gamma}$
will disappear. For $|\lambda_{h_1HH}|< 0.05$,
variation of $R_{\gamma \gamma}$ with $R_{\gamma Z}$ is almost linear
(with slope $\approx 1$)  which
is presented by the line passing through origin shown in plots 
of Fig. \ref{fig2}. 
The scattered plots represent the correlation for other 
values of $\lambda_{h_1HH}$.
For low mass dark matter $(m_H \lesssim m_1/2)$, invisible decay 
channel of $h_1$ remains open and the processes
$h_1\rightarrow \gamma \gamma$ and $h_1\rightarrow  \gamma Z$ suffer 
considerable suppressions.
These result in the correlation between the channels 
$h_1\rightarrow \gamma \gamma$
and $h_1\rightarrow \gamma Z$ to become stronger and $R_{\gamma \gamma}$
{\it vs} $R_{\gamma Z}$ plot shows more linearity 
with increase in $h_2$ mass.
For larger $h_2$ masses, the corresponding charged scalar ($H^{\pm}$)
masses for which $R_{\gamma \gamma,\gamma Z} > 1$, 
tends to increase. Since any increase in $H^{\pm}$ mass will affect the
contribution from charged scalar loop, the decay widths
$\Gamma(h_1\rightarrow \gamma \gamma,\gamma Z)$ or signal strengths 
$R_{\gamma \gamma,\gamma Z}$ are likely to reduce. 
Our numerical results exhibit a positive correlation between 
the signal strengths
$R_{\gamma \gamma}$ and $R_{\gamma Z}$. This is 
an important feature of the model.
Since signal strengths tend to increase with relatively smaller values of
$m_2$, possibility of having a light singlet like scalar
is not excluded. The coupling of $h_2$ with SM sector is suppressed
by a factor $\sin\alpha$ which results in a decrease in the signal
strengths from $h_2$ and makes their observations difficult.
\section{Direct Detection}
\label{S:ddetection}
Within the framework of our model and allowed values of parameter 
region obtained
in Sec.~\ref{S:Analysis}, we calculate spin independent (SI) 
elastic scattering
cross-section for the dark matter candidate in our model 
off a nucleon in the
detector material. We then compare our results with those 
given by various direct
detection experiments and examine the plausibility of our 
model in explaining the
direct detection experimental results.   
The DM candidate in the present model, interacts with SM via 
processes led by Higgs exchange.
The spin-independent elastic scattering cross section 
$\sigma_{\rm{SI}}$ is of the form
\bea
\sigma_{\rm {SI}}\simeq \frac{m_r^2}{\pi}\left(\frac{m_N}{m_H}\right)^2 f^2
\left(\frac{\lambda_{h_1HH}\cos\alpha}{m_1^2}+
\frac{\lambda_{h_2HH}\sin\alpha}{m_2^2} \right)^2,
\label{21}
\eea  
where $m_N$ and $m_H$ are the masses of scattered nucleon and DM respectively, 
$f$ represents the
scattering factor that depends on pion-nucleon cross-section and 
quarks involved in
the process and $m_r=\frac{m_Nm_H}{m_N+m_H}$ is the reduced mass.
In the present framework $f= 0.3$ \cite{hall} is considered. 
The computations of $\sigma_{\rm{SI}}$ for the dark matter candidate 
in the present model 
are carried out with those values of the couplings restricted 
by the experimental value of relic density. 
\begin{figure}[h!]
\centering
\subfigure[]{
\includegraphics[scale= 0.65,angle=0]{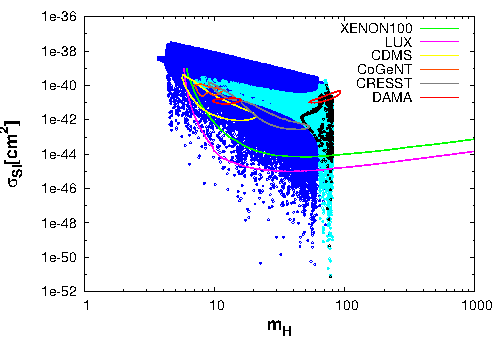}}
\subfigure []{
\includegraphics[scale= 0.65,angle=0]{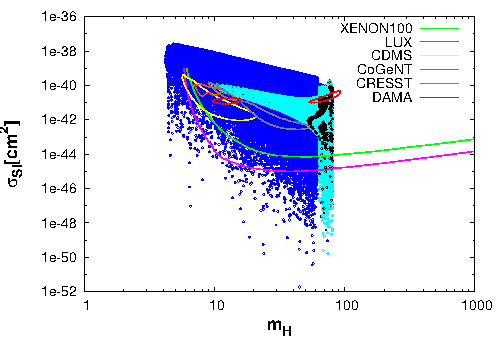}}
\subfigure []{
\includegraphics[scale= 0.65,angle=0]{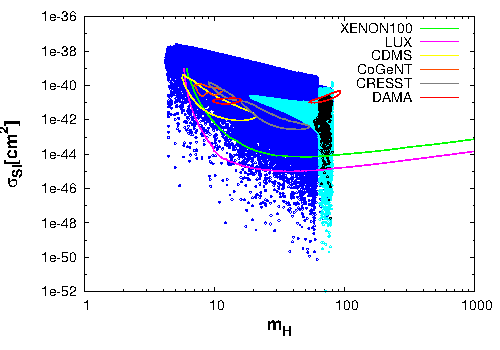}}
\caption{Allowed regions in $m_H-\sigma_{SI}$ plane for $m_2= 140,150~\rm{and}~160$ GeV.}
\label{fig3}
\end{figure}

\begin{figure}[h!]
\centering
\subfigure[]{
\includegraphics[scale= 0.65,angle=0]{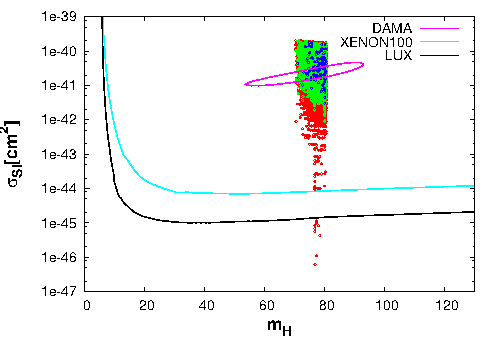}}
\caption{The  $m_H$ vs $\sigma_{SI}$ parameter space for 
$R_{\gamma \gamma} \gtrsim 1$ for $m_2=140-160$ GeV.}
\label{fig4}
\end{figure}
In Fig. \ref{fig3}(a-c), we present the variation of elastic 
scattering cross section
calculated using Eq. \ref{21}, with LIP
dark matter mass ($m_H$)
for three values of $h_2$ masses $m_2 = 140, 150$ and $160$ GeV.
We assume $h_1$ to be SM-like Higgs and restrict the mixing 
angle $\alpha$ such that 
the conditon $\cos \alpha \gtrsim 1/\sqrt{2}$ is satisfied. 
Also shown in Fig. \ref{fig3}(a-c), superimposed on the computed results, 
the bounds on $\sigma_{\rm SI}-$ 
DM mass obtained from DM direct
search experiments such as XENON100, LUX, CDMS, CoGeNT, CRESST.
From Fig. \ref{fig3} one notes that   
in the low mass region, the DM candidate in our model satisfies
bounds obtained from experiments like CoGeNT, CDMS, CRESST.
We further restrict the $\sigma_{\rm{SI}}-m_{H}$ space by identifying in
Fig. \ref{fig3}(a-c) the region for which the CMS limit of $R_{\gamma \gamma}$
($R_{\gamma \gamma}=0.78^{+0.28}_{-0.26}$) is satisfied.
In each of the $\sigma_{\rm{SI}}-m_{H}$ plots of Fig. \ref{fig3}(a-c) the
light blue region satisfies CMS limit of $R_{\gamma \gamma}$ for three
chosen values of $m_2$. Also marked in black are the specific zones 
that correspond 
to the central value of $R_{\gamma \gamma}|_{\rm{CMS}} = 0.78$. It is therfore 
evident 
from Fig. \ref{fig3}(a-c) that imposition of signal
strength ($R_{\gamma \gamma}$) results obtained from LHC, 
further constraints the allowed scattering
cross-section limits  obtained from direct detection experimental results
for the DM candidate in our model. Investigating the region allowed by LUX
and XENON experiments along with other direct dark matter 
experiments such as
CDMS etc., it is evident from Fig. \ref{fig3}(a-c) that our model suggests
a DM candidate within the range $m_1/2 < m_H < m_W$ GeV with scattering cross-section values
$\sim 10^{-44}-10^{-48}$ cm$^2$ with $m_1=125$ GeV, i.e., SM-like scalar.
There are however few negligibly small allowed parameter
space with $\sigma_{\rm SI}$ below $\sim 10^{-48}$ cm${^2}$.
It may also be noticed from Fig. \ref{fig3} that the present model with
all the constraints including $R_{\gamma \gamma}$ condition also partly
agrees with allowed contour given by DAMA experiment. However, DAMA 
contour is ruled out
by recent results from experiments like LUX and XENON100.
Similar procedure has been adopted for restricting the $\sigma_{\rm SI}-m_{H}$
space with $R_{\gamma \gamma}$ limits from ATLAS experiment. 
In Fig. \ref{fig4}, the region shown in red corresponds to the region 
satisfying $R_{\gamma \gamma} \gtrsim 1$ with 
mass of $h_2$ varied from $140$ GeV to $160$ GeV. Also shown in Fig. \ref{fig4},
the scattered region in green (blue) represents the signal strength 
$R_{\gamma \gamma}=1.65^{+0.34}_{-0.30}$ ($R_{\gamma \gamma} =1.65$) 
as obtained from ATLAS experiment respectively.
Fig. \ref{fig4} shows that the part of the region constrained by
ATLAS result is more stringent than that for CMS case and appears to
satisfy only a part of DAMA allowed contour.
There is however a negiligibly 
small allowed region satisfying the domain constrained by LUX or XENON100
expreiments. Similar to the case for $R_{\gamma \gamma}$ limit from CMS, here
too, the allowed zone lies in the range around $m_{H} = 70$ GeV.
Hence, in the present model $H$
can serve as a potential dark matter candidate and future experiments 
with higher sensitivity like XENON1T \cite{xe1T}, SuperCDMS \cite{scdms} 
etc. are expected
to constrain or rule out the viability of this model. 
\begin{figure}[h!]
\centering
\subfigure[]{
\includegraphics[scale= 0.66,angle=0]{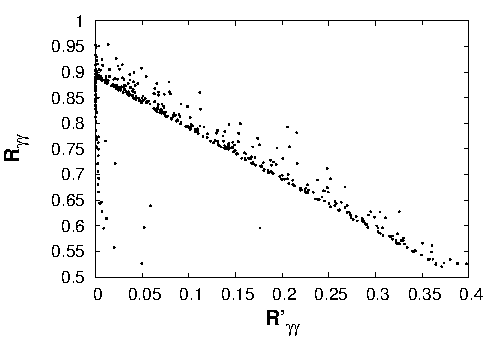}}
\subfigure []{
\includegraphics[scale= 0.66,angle=0]{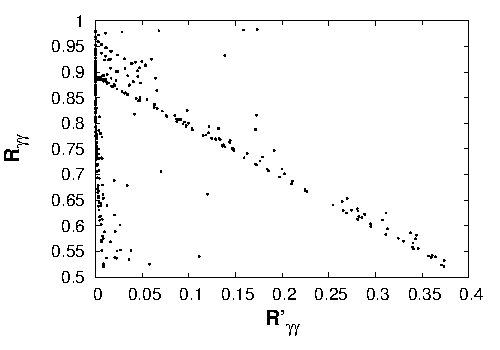}}
\subfigure []{
\includegraphics[scale= 0.66,angle=0]{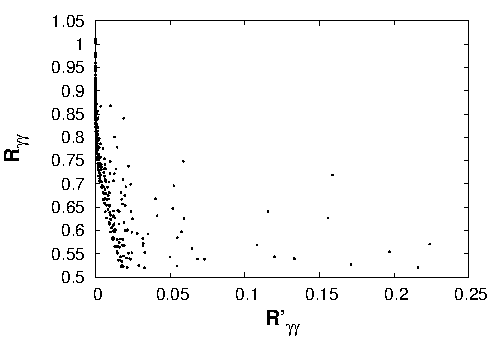}}
\caption{Allowed regions in $R_{\gamma \gamma}-R'_{\gamma \gamma}$ plane for $m_2= 140,150~\rm{and}~160$ GeV.}
\label{fig5}
\end{figure}
In the present model we so far adopt the consideration that $h_1$ plays the role of 
SM Higgs and hence in our discussion we consider $h_1\rightarrow \gamma \gamma$
for constraining our parameter space. The model considered in this work
also provides us with a second scalar namely $h_2$. Since LHC has not 
yet observed a second scalar, it is likely that the other 
scalar $h_2$ is very weakly
coupled to SM sector so that the corresponding branching 
ratios (signal strengths)
are small. This may be justified in the present scenario if in case the 
$h_2\rightarrow \gamma \gamma$ branching ratio or signal 
strength ($R'_{\gamma \gamma}$)
is very small compared to that for $h_1$. Needless to mention that the couplings
required to compute $R_{\gamma \gamma}$ and $R'_{\gamma \gamma}$ are restricted
by dark matter constraints. One also has to verify whether the process
$R'_{\gamma \gamma}$ can play significant role in restricting the dark matter 
model parameter space in the present framework. We address 
these issues by computing
$R'_{\gamma \gamma}$ values and comparing them with $R_{\gamma \gamma}$
\footnote{Since $R'_{\gamma \gamma}$ and
$R'_{\gamma Z}$ are correlated, any suppression in
$h_2 \rightarrow \gamma \gamma$ will be followed by similar effects 
in $h_2 \rightarrow \gamma Z$.}. 
%Hence we compare values $R'_{\gamma \gamma}$ with $R_{\gamma \gamma}$ only.}.
The computation of 
$R_{\gamma \gamma}$ and $R'_{\gamma \gamma}$ initially involves the dark 
matter model parameter space that yields the dark matter relic density 
in agreement with PLANCK data as also the stringent direct detection 
cross-section bound obtained from LUX. 
$R_{\gamma \gamma}$ values thus obtained are not found to satisfy the 
experimental range given by ATLAS experiment. 
The resulting $R_{\gamma \gamma} - R'_{\gamma \gamma}$ is further restricted
for those values of $R_{\gamma \gamma}$  which are within the 
limit of $R_{\gamma \gamma}|_{\rm{CMS}}$ given by CMS experiment.  
The region with black 
scattered points in Fig. \ref{fig5}(a-c) corresponds to the
$R_{\gamma \gamma}-R'_{\gamma \gamma}$ space 
consistent with the model parameters that are allowed by DM relic 
density obtained from PLANCK, direct detection experiment
bound from LUX and $R_{\gamma \gamma}|_{\rm{CMS}}$
for three different values of $m_2~(140,~150,~{\rm and}~160~{\rm GeV}$). 
Fig. \ref{fig5}(a-c) reveals that $R'_{\gamma \gamma} \leq 0.4$ for 
$m_2 = 140,~150$ GeV whereas for $m_2 = 160$ GeV it is even less 
($R'_{\gamma \gamma} < 0.25$) for $R_{\gamma \gamma}$ values compatible with
CMS results. In Table \ref{tab1} we further
demonstrate that within the framework of our proposed model for 
LIP dark matter, $R'_{\gamma \gamma}$ is indeed small
compared to $R_{\gamma \gamma}$. We tabulate the values of both $R_{\gamma \gamma}$
and $R'_{\gamma \gamma}$ for some chosen values of LIP dark matter mass $m_H$.
These numerical values are obtained from the computational results consistent
with LUX direct DM search bound. Also given in Table \ref{tab1} the corresponding
mixing angles $\alpha$ between $h_1$ and $h_2$, the scalar masses $m_H^{\pm}$,
$h_2$ to di-photon branching ratio
and the scattering cross-section $\sigma_{\rm{SI}}$ for three different values
of $m_2$ considered in the work. It is also evident from Table \ref{tab1} that
$R_{\gamma \gamma}>>R'_{\gamma \gamma}$ and mixing angles corresponding to respective values 
are small.
In fact for some cases such as for $m_H=70.13$ GeV ($m_2=140$ GeV) 
$R_{\gamma \gamma}=0.889$ whereas $R'_{\gamma \gamma}\sim 10^{-3}$ and $\alpha$ is
as small as 2. This demonstrates that the scalar $h_1$ in Eq. \ref{4} is mostly
dominated by SM-like Higgs component and the major component in the other scalar
is the real scalar singlet $s$ of the proposed model.
Table \ref{tab1} also exhibits that the signal strength $R'_{\gamma \gamma}$ for
$h_2\rightarrow \gamma \gamma$ channel is negligibly weak.
\begin{table}
  
\begin{center}
\vskip 0.5 cm
\begin{tabular}{|c|c|c|c|c|c|c|c|}
\hline
         &        &                &            &                     &                      &                                  &      \\
$m_2$    & $m_H$  & $m_{H^{\pm}}$  &  $\alpha$  & $R_{\gamma \gamma}$ & $R'_{\gamma \gamma}$ &$Br(h_2\rightarrow \gamma \gamma)$&$\sigma_{SI}$      \\ 
 in GeV  & in GeV &    in GeV      &            &                     &                      &                                  &  in $\rm{cm^2}$    \\
\hline  
         & 70.13  &    231.00      &    02      &        0.889        &        1.13e-03      &            1.835e-03             &  1.360e-47    \\ 
  
 140.00  & 67.45  &    151.00      &    12      &        0.856        &        4.48e-05      &            2.038e-06             &  7.405e-48    \\ 

         & 78.40  &    302.50      &    07      &        0.873         &       1.70e-02      &            2.256e-03             &  2.046e-46    \\
\hline
         & 64.05  &    181.00      &    20      &        0.786        &        3.50e-03      &            4.220e-05             &  6.386e-46    \\

 150.0   & 79.57  &    242.50      &    04      &        0.889        &        2.63e-03      &            7.613e-04             &  5.749e-47    \\ 

         & 80.29  &    138.00      &    10      &        0.875        &        3.67e-02      &            1.715e-03             &  8.437e-48     \\
\hline
         & 66.67  &    206.50      &    13      &        0.844        &        5.67e-05      &            6.217e-07             &  6.961e-50     \\

 160.0   & 74.49  &    136.50      &    06      &        0.900        &        1.80e-07      &            9.165e-09             &  1.491e-47     \\
         
         & 77.86  &    311.50      &    16      &        0.821        &        1.71e-03      &            1.255e-05             &  7.371e-49     \\
\hline
\end{tabular}
\end{center}
\caption{Benchmark points satisfying observed DM relic density obtained from PLANCK data and
direct detection cross-section reported by LUX results for three different choices of $h_2$ mass.}
\label{tab1}
\end{table}

\section{Summary}
\label{S:summary}
In this paper we have proposed a model for dark matter where we consider
an extended two Higgs doublet model with an additional
singlet scalar. The
DM candidate follows by setting one of the Higgs doublet to be identical with an inert
Higgs doublet imposing a $Z_2$ symmetry   
on the potential. This ensures the DM candidate that 
follows from the added inert doublet is stable. The inert doublet does not generate
any VEV and hence cannot couple to Standard Model fermions directly. The scalar singlet, having no such 
discrete symmetry aquires a non zero VEV and mixes up with SM Higgs.
The unknown couplings appearing in the model, 
which are basically the model parameters, are restricted with theoretical and experimental 
bounds.
The mixing of the SM and the singlet
scalar gives rise to two sclar states namely $h_1$ and $h_2$. For small mixing $h_1$ behaves as 
the SM Higgs and $h_2$ as the added scalar. 
We extensively explored the scalar sector
of the model and studied the signal streghts $R_{\gamma \gamma}$ and $R_{\gamma Z}$ for the SM-
like Higgs $(h_1)$ in the model. 
The range and region of enhancement depens on the mass of the singlet like scalar $h_2$.
Appreciable enhancement of signals depends on $h_2$ mass
which occurs near the Higgs resonance. 
Increase in signal strengths is not allowed for heavier values of $h_2$ mass.
Enhancement of signals are forbidden when the invisible decay channel
remains open. The extent of enhancement depends on the charged scalar mass and occurs only when the Higgs-charged scalar
coupling $\lambda_{h_1H^+H^-} < 0$.
We first restrict our parameter space by calculating the relic density of LIP dark matter
in the framework of our model. Using the resultant paramter space obtained from relic density bounds
we evalute
the signal strengths $R_{\gamma \gamma}$ and $R_{\gamma Z}$ for different dark matter mass.
We then restrict the parameter space by calculating the spin independent
scattering cross-section and comparing it with the existing limits from ongoing
direct detection experiments like CDMS, CoGeNT, DAMA, XENON100, LUX etc.
Employing additional constraints by requiring that $R_{\gamma \gamma}$ and 
$R_{\gamma Z} $ will satisfy the CMS bounds and ATLAS bounds, we see that 
the present model not only provides a good DM candidate in middle mass region 
consistent with LUX and XENON100 bounds. The possibility that $R_{\gamma \gamma}
(>1.0)$ in the present framework does not seem to be 
favoured by LUX and XENON100
data. However, DAMA results appear to favour $R_{\gamma \gamma} \geq 1$.
Therefore, we conclude that under the present framework, Inert Doublet Model
with additional scalar singlet provide a viable DM candidate
with mass range $m_1/2 < m_H < m_W$ GeV that not only is consistent 
with the direct detection experimental bounds and PLANCK results 
for relic density but also in agreement with the Higgs search results  
of LHC.
A singlet like scalar that couples weakly with SM Higgs may also exist  
that could enrich the Higgs sector and may be probed in future 
collider experiments.

{\bf Acknowledgments} : A.D.B. would like to thank A. Biswas, D. Das and 
K.P. Modak
for useful discussions.

%\appendix
%\appendixpage
%\section{Appendix}
%\appendix
%\label{App}
\vskip 2mm
\noindent {\bf Appendix A}
\vskip 2mm

\noindent In Section 3.3 we have derived the decay 
widths $h_1\rightarrow \gamma \gamma$
and $h_1\rightarrow \gamma Z$ in terms of the loop factors
$F_{1/2},~F_1,~F_0,~F_{1/2}',~F_1'$ and $I_1$ respectively.
Factors $F_{1/2},~F_1,~F_0$, for the measurement of 
$h_1\rightarrow \gamma \gamma$
decay width can be written as \cite{higgshunter,djouadi1,djouadi2}   
\bea
F_{1/2}(\tau)&=&2\tau[1+(1-\tau)f(\tau)],\nonumber\\
F_1(\tau)&=&-[2+3\tau+3\tau(2-\tau)f(\tau)],\nonumber\\
F_0(\tau)&=&-\tau[1-\tau f(\tau)],\nonumber
\eea
and
\bea
f(\tau)=\left\{ \begin{array}{ll}
\arcsin^2\left(\frac{1}{ \sqrt{\tau} }\right) & \rm{for}~~~~\tau \geq 1,\\
-\frac{1}{4}\left[\log\left(\frac{1+\sqrt{1-\tau}}{1-\sqrt{1-\tau}}\right)-i\pi\right]^2  & \rm{for}~~~~ \tau<1.
\end{array} \right.
\nonumber
\eea
Loop factors for the decay $h_1 \rightarrow \gamma Z$ are expressed following Refs.
\cite{higgshunter,djouadi1,djouadi2} 
\bea
F_{1/2}'(\tau,\lambda)&=&I_1(\tau,\lambda)-I_2(\tau,\lambda),\nonumber\\
F_1'(\tau,\lambda)&=&c_W\left\{4\left(3-\frac{s^2_W}{c^2_W}\right)I_2(\tau,\lambda)+
\left[\left(1+\frac{2}{\tau}\right)\frac{s^2_W}{c^2_W}-\left(5+\frac{2}{\tau}\right)\right]I_1(\tau,\lambda)\right\},\nonumber
\eea
where
\bea
I_1(a,b)&=&\frac{ab}{2(a-b)}+\frac{a^2b^2}{2(a-b)^2}\left[f(a)-f(b)\right]+\frac{a^2b}{(a-b)^2}\left[g(a)-g(b)\right],\nonumber\\
I_2(a,b)&=&-\frac{ab}{2(a-b)}\left[f(a)-f(b)\right].\nonumber
\eea
Expressions of $g(\tau/\lambda)$ is given by
\bea
g(\tau)=\left\{ \begin{array}{ll}
\sqrt{\tau-1}\arcsin\sqrt{\frac{1}{\tau}}& \rm{for}~~~~\tau \geq 1,\nonumber\\
\frac{\sqrt{1-\tau}}{2}\left(\log\frac{1+\sqrt{1-\tau}}{1-\sqrt{1-\tau}}-i\pi\right)&\rm{for}~~~~\tau<1.
\end{array} \right.
\eea

\end{document}